\documentclass[3p]{elsarticle}

\usepackage{amsmath}
\usepackage{lineno,hyperref}
\modulolinenumbers[5]

\journal{Journal of Computational Physics}

\usepackage{lipsum}

\makeatletter
\def\ps@pprintTitle{%
 \let\@oddhead\@empty
 \let\@evenhead\@empty
 \def\@oddfoot{\footnotesize\itshape
       Published in the \ifx\@journal\@empty Elsevier
       \else\@journal\fi\hfill} %\today}%
 \let\@evenfoot\@oddfoot}
\makeatother

\usepackage{amsfonts}
\usepackage{graphicx}
\usepackage{epstopdf}
\usepackage{algorithmic}
\usepackage{soul}
\usepackage{float}
\usepackage{multirow}
\usepackage{amsbsy}
\usepackage{amssymb}
\usepackage{amsmath} 
\usepackage{algorithmic}
\usepackage{algorithm}
\usepackage{mathtools}
\usepackage{subcaption}
\usepackage{soul}
\usepackage{color}
\definecolor{lightblue}{rgb}{.80,.90,1}

\usepackage{lineno}
\usepackage{ulem}
\usepackage{tikz}
\usetikzlibrary{shapes,arrows}

\tikzstyle{decision} = [diamond, draw, fill=blue!20,
    text width=4.5em, text badly centered, inner sep=0pt]
\tikzstyle{block} = [rectangle, draw, fill=blue!20,
    text width=6em, text centered, rounded corners, minimum height=4em]
\tikzstyle{line} = [draw, -latex']
\tikzstyle{cloud} = [draw, ellipse,fill=red!20, node distance=3cm,
    minimum height=2em]
\newdefinition{rmk}{Remark}

\begin{document}

\begin{frontmatter}

\title{Multifidelity Modeling for Physics-Informed Neural Networks (PINNs)}
%\tnotetext[mytitlenote]{Fully documented templates are available in the elsarticle package on %\href{http://www.ctan.org/tex-archive/macros/latex/contrib/elsarticle}{CTAN}.}

\author[SCI]{Michael Penwarden}
\ead{mpenwarden@sci.utah.edu.}
\author[SOC]{Shandian Zhe}
\ead{zhe@cs.utah.edu}
\author[MATH]{Akil Narayan}
\ead{akil@sci.utah.edu}
\author[SCI]{Robert M. Kirby}
\ead{kirby@cs.utah.edu}

\address[SCI]{School of Computing and Scientific Computing and Imaging Institute, University of Utah, Salt Lake City, UT}
\address[SOC]{School of Computing, University of Utah, Salt Lake City, UT}
\address[MATH]{Department of Mathematics and Scientific Computing and Imaging Institute, University of Utah, Salt Lake City, UT}

\begin{abstract}
Multifidelity simulation methodologies are often used in an attempt to judiciously combine low-fidelity and high-fidelity
simulation results in an accuracy-increasing, cost-saving way.  Candidates for this approach are simulation methodologies
for which there are fidelity differences connected with significant computational cost differences.  Physics-informed
Neural Networks (PINNs) are candidates  for these types of approaches due to the significant difference in training times 
required when different fidelities (expressed in terms of architecture width and depth as well as optimization criteria) are employed.  In this paper, we 
propose a particular multifidelity approach applied to PINNs that exploits low-rank structure.  We demonstrate that width, depth, and optimization criteria can
be used as parameters related to model fidelity and show numerical justification of cost differences in training due to
fidelity parameter choices.   We test our multifidelity scheme on various canonical forward PDE models
that have been presented in the emerging PINNs literature.

\end{abstract}


\end{frontmatter}

%\linenumbers

%
\section{Introduction}
\label{sec:introduction}

Engineering is replete with situations in which both low-fidelity (even ``back of the envelope") models and high-fidelity models are available
to aid in decision-making.  It is often the case that the discrepancy between low-fidelity and high-fidelity is associated with a corresponding 
difference in computational cost:  the low-fidelity simulation is far cheaper to compute than a high-fidelity simulation.  These situations
have motivated extensive research into multifidelity methods:  those methods that attempt to judiciously combine low-fidelity and high-fidelity
simulation results in an accuracy-increasing, cost-saving way.  A comprehensive review of recent advances and works in the area of multifidelity methods can be 
found in the manuscript by Peherstorfer et al.~\cite{peherstorfer2018survey}.

Multifidelity construction of surrogate models for a diverse range of systems have been successfully implemented and presented in the literature. In many cases a well-established continuous relationship with respect to a discretization parameter exists, describing the convergence of the low-fidelity model to high-fidelity one. Hence, the notion of ``fidelity" often represents a discrepancy of the low-fidelity model relative to the high-fidelity one, and accounts for amount of discretization coarsening, geometrical simplification, and underlying physical model complexity. Time-step size with theoretical guarantees
used for canonical ODEs \cite{keshavarzzadeh19SIAM}, time-step size
in molecular dynamics simulation~\cite{razi2018MD}, quadratic frequency modulation in frequency-modulated trigonometric functions~\cite{narayan2014stochastic}, finite element mesh size in acoustic horn problems~\cite{zhu2014computational}, finite element mesh size in topological optimization problems \cite{Keshavarzzadeh19IJNME,Keshavarzzadeh21},
finite volume discretization for heat driven cavity flow~\cite{hampton2018practical}, aerodynamic model simplification in the parametric study of NACA airfoils~\cite{skinner2017evaluation} and amount of coarsening of the Eulerian and Lagrangian resolutions for the study of irradiated particle-laden turbulence~\cite{jofre2018multi} are among the examples of fidelity parameters used in the literature for the purpose of multifidelity construction of surrogate models.  In all these cases, both high- and low-fidelity models attempt to model the same problem, and a direct relationship between the two can be perceived; the difference between the models in these cases is often a numerical discretization parameter that is chosen differently so that the low-fidelity model is less expensive, but also less accurate. Thus, in most of these cases the convergence of the low-fidelity model to the high-fidelity model can be proven analytically under refinement of discretization parameters and often there is a hierarchical connection between low- and high-fidelity models. For the cases of geometrical and physical simplifications, e.g., the composite beam example in \cite{hampton2018practical}, the low-fidelity model is indeed a simplified surrogate of the high-fidelity model, the latter of which includes more assumptions about the underlying physics of the system.  Furthermore, one can even apply such methods when the fidelity is represented by the difference in quantized model hierarchies within discrete systems \cite{RaziDiscrete19}.

It is against this backdrop that we consider the extension of the low-rank parametric multifidelity approach of \cite{narayan2014stochastic,zhu2014computational} to Physics-informed Neural Networks (PINNs) \cite{raissi2019physics,raissi2017physicsI,raissi2017physicsII}.  PINNs represent a new ``meshfree" discretization methodology built upon deep neural networks (DNNs), and capitalize on machine learning technologies such as automatic backward differentiation and stochastic optimization \cite{DeepLearning}.  The marriage of computational modeling and machine learning is predicted to transform the way we do science, engineering and clinical practice \cite{Alber2019}.

In this paper, we adapt a multifidelity approach for parametric problems to PINNs, using the width and depth of the network architecture (for fixed activation functions) as well as optimization criteria as the means to determine fidelity levels.  We provide theoretical discussions and experiments to motivate our width, depth and optimization criteria choices and their connection with fidelity.  We also discuss some possible pitfalls of this connection.  In regards to the connection between fidelity and computational cost:  As the width and depth of a DNN is increased, the training time may increase significantly \cite{DeepLearning}. Hence we posit that PINNs is an admissible, if not ideal, candidate for multifidelity approaches.  

The paper is organized as follows:  In Section \ref{sec:multifidelity}, we provide an overview of our low-rank multifidelity approach.  In Section \ref{sec:pinns}, we first review the original PINNs collocation approach and provide a brief summary of current and ongoing PINNs efforts within the field.  Although we focus on the application of our multifidelity approach to the collocation version of PINNs, nothing precludes the extension of our work to other PINNs variants upon appropriate evaluation and minor modifications.  In Section \ref{sec:results},  we present our multifidelity PINNs approach applied to forward problems that have been presented in the emerging PINNs literature.  Furthermore, we discuss the limitations and assumptions within our approach with present open theoretical and methodological challenges to the PINNs and machine learning communities.  We conclude in Section \ref{sec:summary} with a summary of our work and a discussion of current challenges and potential future avenues of inquiry and expansion of the concepts presented in this work.

\section{Overview of our Low-rank Multifidelity Approach}
\label{sec:multifidelity}

\newcommand{\R}{\mathbb{R}}
Consider a low-fidelity and a high-fidelity model denoted, respectively, as,
\begin{align}\label{eq:gLH-def}
  g_L: D &\rightarrow \R^m, & g_H: D &\rightarrow \R^M,
\end{align}
where $D \subset \R^d$ is a $d$-dimensional parameter space. In the context of parameterized partial differential equations (PDEs), we view $g_L$ and $g_H$ as solvers or emulators, mapping a common parameter space $D$ to separate output spaces of different dimensions (e.g., as in coarse/fine or multiscale solvers). We make no assumptions, at this stage, on how $m$ and $M$ are related, but typically $m \ll M$. For a given parameter $p \in D$, we also make no explicit assumption about the physical meaning or interpretation of the model responses $g_L(p)$ versus $g_H(p)$; in particular we do not make formal assumptions ensuring $g_L(p) \approx g_H(p)$.  In Section \ref{sec:pinns}, we will provide some insights on how this setup can be realized in a PINNs framework.

We are interested in developing a multifidelity framework for PINNs, where $g_L$ and $g_H$ aim to model the same high-level system, but with different levels of accuracy or fidelity. In particular we assume the ordered hierarchy that $g_H$ is a more expensive, but also more trusted predictor compared to $g_L$. For PINNs, we will use the width and depth of the network as well as optimization criteria, which are surrogates for expressivity and training optimality respectively, to define fidelity.  In order to tackle this situation, our proposal is to use the methodology from \cite{narayan2014stochastic}, the use of which has subsequently been refined \cite{zhu2014computational,keshavarzzadeh19SIAM,hampton_practical_2018}. This procedure is relatively simple --  containing the following high-level steps: 
\begin{enumerate}[\indent {}]
  \item {\bf Step 1}: Discretize parameter space $D$ with $K \gg 1$ samples $D_L \coloneqq \{p_1, \ldots, p_K\}$.
  \item {\bf Step 2}: Evaluate $g_L$ on the $K$ points $D_L$, and use this to identify a subset of $k \ll K$ points $D_H = \{p_{i_1}, \ldots, p_{i_k}\} \subset D_L$ -- the so-called ``important" points in \cite{narayan2014stochastic}. 
  \item {\bf Step 3}: Evaluate $g_H$ at the $k$ points in $D_H$.
  \item {\bf Step 4}: Construct a multifidelity emulator using stored low and high fidelity information at the ``important" points.
\end{enumerate}
Once these steps are completed, one has an emulator for the high-fidelity model constructed with the cost of dense sampling of the low-fidelity model and very few ($k$) high-fidelity
samples. The evaluation of the emulator at any given parameter $p$ costs a single low-fidelity model evaluation, but is an emulator for the high-fidelity model $g_H$. This is a low-rank procedure because the selection of points in Step 2, and the constructions in Step 4, exploits low-rank structure in certain matrices. We provide the details now. 

The choice of $D_L$ is problem-dependent; for example, $K$ values uniformly sampled at random from the parameter space $D$ is often used. Once the low-fidelity model $g_L$ is evaluated at every point in $D_L$, a $K \times K$ Gram matrix $G_L$ is constructed with the following entries,
\begin{eqnarray}
  (G_L)_{i,j} &=& \left\langle g_L(p_i), g_L(p_j) \right\rangle  \mbox{\ \ }i, j = 1, \ldots, K,
\label{eq:E6}
\end{eqnarray}
where $\left\langle \cdot, \cdot \right\rangle$ is often chosen as the standard Euclidean inner product on the low-fidelity output space $\R^m$, corresponding to a \textit{linear} kernel on two input features.
Alternative inner product definitions, or choices of kernels, can be used depending on the properties of the Gramian that are desired; 
Razi et al. studied the impact of different kernels in comparison to the standard linear kernel above \cite{Razi21}.

Note that the above is identical to formation of $G_L$ via
\begin{align}\label{eq:VL-def}
  G_L &= V_L^T V_L, & V_L &\coloneqq \left(\begin{array}{cccc} g_L(p_1) & g_L(p_2) & \cdots & g_L(p_K) \end{array}\right) \in \R^{m \times K}.
\end{align}
The procedure by which the $k$ ``important" points $D_H$ are identified from $D_L$ utilizes the matrix $G_L$ (or, equivalently, $V_L$). Selecting ``important" rows/columns from a matrix is a matrix subset selection problem that is related to low-rank approximation.

There are a number of procedures that provides selection tools for most important $k$ indices from a Gramian matrix. Among the most commonly used selection approaches in the literature include linear algebraic strategies, such as pivots chosen by a column-pivoted QR decomposition of $V_L$~\cite{narayan2014stochastic}, or equivalently the pivoted Cholesky decomposition of the Gram matrix $G_L$~\cite{zhu2014computational}, the LU factorization~\cite{anderson2017efficient}; statistical strategies such as leverage score sampling methods~\cite{perry2016augmented, Perry19}; and sparsity-promoting group matching methods~\cite{lozano2011group,Perry19}. In this paper, we choose to use pivots identified from a pivoted Cholesky decomposition for this purpose: the procedure is easy to understand, readily implemented, available on many computational platforms, computationally efficient, and in our experience performs competitively with alternative methods \cite{Perry19}. The Cholesky approach forms the following pivoted Cholesky decomposition
\begin{linenomath}\begin{align*}
  P^T G_L P = R^T R, \; R = \begin{bmatrix}
R_{11} & R_{12} \\
0 & 0
\end{bmatrix}
\end{align*}\end{linenomath}
where $R_{11}$ is a square matrix, and $P \in \R^{K \times K}$ is a permutation matrix whose entries identify column pivot indices in the decomposition process. We label these indices as $i_1, \ldots, i_K$, which are a permutation of the set $(1, \ldots, K)$.

It is well-known that the pivots chosen in this way are equivalent to those produced from a column-pivoted QR decomposition of $V_L$ \cite{golub_matrix_1996}. We have framed this discussion in the context of storing the full matrix $G_L$ and computing the superfluous $R_{11}$ and $R_{12}$ factors, but one can construct algorithms that need not store the potentially large matrix $G_L$, and compute only the pivots in $P$; see, e.g., \cite{narayan2014stochastic}.

The multifidelity procedure chooses $D_H$ from the pivots identified above:
\begin{linenomath}\begin{align*}
  D_H = \left\{ p_{i_1}, \ldots, p_{i_k} \right\}.
\end{align*}\end{linenomath}
The number $k$, representing the number of high-fidelity model evaluations that must be run, is chosen based on the available computational budget for the high-fidelity model. Because of the expense of the high-fidelity model, we frequently have that $k$ is $\mathcal{O}(10)$ in practice. Next, we compute the high-fidelity model $g_H\left(p_{i_l}\right)$ at these points; in situations of interest, this $k$-fold query of the high-fidelity model is typically the most expensive step of the procedure. Finally, the multifidelity approximation $\widetilde{g}_H$ can be constructed from these simulations:
\begin{linenomath}\begin{align}\label{eq:mf-surrogate}
  g_H(p) \approx \widetilde{g}_H(p) \coloneqq \sum_{l=1}^k g_H\left(p_{i_l}\right) c_l(p),
\end{align}\end{linenomath}
where $\{c_l(p)\}_{l=1}^k$ are coefficients computed via a least-squares projection from the low-fidelity model:
\begin{eqnarray}
  \left(\begin{array}{ccc} 
      (G_L)_{i_1, i_1}  & \cdots & (G_L)_{i_1,i_k} \\
      \vdots                    & \ddots & \vdots \\
  (G_L)_{i_k, i_1}  & \cdots & (G_L)_{i_k,i_k}\end{array}\right)
  \left(\begin{array}{c} c_1(p) \\ \vdots \\ c_k(p) \end{array}\right)
    = \left( \begin{array}{c} \left\langle g_L(p), g_L(p_1) \right\rangle \\
                                     
                                     \vdots \\
    \left\langle g_L(p), g_L(p_k) \right\rangle \end{array}\right). \nonumber \\
\label{eq:E9}
\end{eqnarray}
The coefficients $c_j(p)$ correspond to weights in a least-squares approximation of $g_L(p)$ using the basis $\{g_L(p_{i_j})\}_{j=1}^k$. The multifidelity technique therefore computes least-squares coefficients from the low-fidelity model, and uses these coefficients in the high-fidelity prediction. The coefficients $\{c_l(p)\}_{l=1}^k$ can also be computed directly from low-fidelity snapshots as,
\begin{linenomath}\begin{align*}
  c(p) &= V^\dagger_K g_L(p) \in \R^K, & V_K &\coloneqq \left( g_L(p_{i_1}), \ldots, g_L(p_{i_K})\right) \in \R^{m \times K},
\end{align*}\end{linenomath}
where $A^\dagger$ is the Moore-Penrose pseudoinverse of $A$ \cite{Harville}. An alternative to this approach would use least-squares to project the full low-fidelity ensemble onto the $k$ important points, involving an $(m \times k)$ version of \eqref{eq:E9}.

Because \eqref{eq:E9} is a square system, the approximation $\widetilde{g}_H$ is interpolatory at the important points:  $\widetilde{g}_H(p) = g_H(p)$ for every $p \in D_H$. The evaluation of $\widetilde{g}_H$ can be accomplished with a single evaluation of the low-fidelity model $g_L$, which is required to form the right-hand-side of \eqref{eq:E9}.

The overall accuracy of this approach, i.e., the efficacy of \eqref{eq:mf-surrogate}, depends on the discrepancy between $G_L$ and $G_H$, but the actual model responses $g_L$ and $g_H$ need not have similar outputs \cite{narayan2014stochastic}. Thus snapshot proximity, i.e.,  $g_L(p) \approx g_H(p)$, is \textit{not} necessary for success of this procedure. Instead, we require a more subtle condition that the parameter variation of $g_L$ is similar to that of $g_H$. See \cite{narayan2014stochastic} for details and theoretical analysis. A practical procedure to estimate the error of this approach is provided in \cite{hampton2018practical}, and in Section \ref{sssec:theory} we provide futher discussion regarding the accuracy of \eqref{eq:mf-surrogate}.

\begin{rmk}
  In this paper, we provide bi-fidelity results as they give the starkest difference in levels. This is in part due to the well-known accuracy limit of PINNs. However, there is no reason more levels cannot be added. One would simply use the bi-fidelity method shown here in sequence to go through the intermediary fidelities. See \mbox{\cite{narayan2014stochastic}} for a description of the multi-fidelity procedure with more than two levels.
\end{rmk}

\section{Physics-Informed Neural Networks (PINNs)}
\label{sec:pinns}

In this section, we first present a review of Physics-Informed Neural Networks (PINNs), with an emphasis on the original collocation PINNs approach which we use in this work.
We also provide a brief summary of current and ongoing PINNs efforts within the field -- many if not all of which might benefit from the multifidelity approach presented herein.
We then present the application of our proposed multifidelity approach to PINNs.  We first provide a summary of the theoretical considerations upon which our work 
is built, and subsequently we provide a summary of the implementation considerations that are required.

%%%%%%%%%%%%%%%%%%%%%%%%%%%%%%
\subsection{Review of Physics-Informed Neural Networks}\label{ssec:pinns-review}

 Physics-Informed Neural Networks (PINNs) were originally proposed by Karniadakis and co-workers \cite{raissi2019physics,raissi2017physicsI,raissi2017physicsII} 
 as a neural network based alternative to traditional PDE discretizations. In the original PINNs work, when presented with a PDE specified 
 over a domain $\Omega$ with boundary conditions on $\partial \Omega$ and initial conditions at $t=0$ (in the case of time-dependent PDEs),
 the solution is computed (i.e. the differential operator is satisfied) 
 as in other mesh-free methods like RBF-FD \cite{ShankarWFK1,ShankarWFK2} at a collection of collocation points.  
 First, we re-write our PDE system in residual form as $R(u) = f - \mathcal{N}(u)$ for an arbitrary differential operator 
 $\mathcal{N}(u)$ which may be a function of both space and time. 
 The PINNs formulation is expressed as follows:  Let $\tilde{u}(x,t;\underbar{w})$ denote a neural network predictor with inputs $(x,t)$ and parameters/weights $\underbar{w}$ that are the degrees of freedom of the network collected from its associated width and depth. In this paper, we assume $\underbar{w}$ is flattened and represented as a vector. Throughout the discussion, the activation function of the network is given and fixed. The network is trained based finding the weights
$\underbar{w}$ that minimize the loss function:

\begin{equation}
MSE = MSE_u + MSE_R
\end{equation}

\noindent where

\begin{subequations}\label{eq:MSE}
\begin{eqnarray}
MSE_u &=& \frac{1}{N_u} \sum_{i=1}^{N_u} \| \tilde{u}(x_u^i,t_u^i;\underbar{w}) - u^i \|^2  \\
  MSE_R &=& \frac{1}{N_R} \sum_{i=1}^{N_R} \| R(\tilde{u}(x_R^i,t_R^i)) \|^2 \,\,\, 
\end{eqnarray}
\end{subequations}

\noindent where $\{x_u^i,t_u^i,u^i\}_{i=1}^{N_u}$ denote the initial and boundary training data on $u(x,t)$ 
and $\{x^i_R,t^i_R\}_{i=1}^{N_R}$ specify the collocation points for evaluation of the collocating residual term $R(\tilde{u})$.  
The loss $MSE_u$ corresponds to the initial and boundary data while $MSE_R$ enforces the structure imposed by the
differential operator at a finite set of collocation points. 

Beyond the initial collocation version of PINNs expressed above, Karniadakis and collaborators have extended these methods to 
conservative PINNs (cPINNs)  \cite{JAGTAP2020113028}, variational PINNs (vPINNS) \cite{kharazmi2019variational}, parareal PINNs (pPINNs) \cite{MENG2020113250}, stochastic PINNs (sPINNs) \cite{ZHANG2019108850}, fractional PINNs (fPINNs) \cite{pang2018fpinns}, LesPINNs (LES PINNs) \cite{PhysRevFluids.4.034602}, non-local PINNs (nPINNs) \cite{pang2020npinns} and eXtended PINNs (xPINNs) \cite{JagtapK}.  

In this work, we will focus on application of the original collocation PINNs approach; however, the work presented herein can be applied
to many if not all of these variants.  

\subsection{Expressivity of neural networks}\label{ssec:expressivity}
PINNs are a special type of neural network (NN) formed from compositions of affine maps and componentwise activation functions where \textit{depth} is the number of compositions and \textit{width} is the number of intermediate outputs in each layer. A PINN regressor $\tilde{u}$ attempts to emulate the map $(x,t) \mapsto u(x,t)$. Solutions to nonlinear PDEs over large spatial and/or time scales result in very complex behavior of $u$ as a function of the inputs $(x,t)$, and the \textit{expressivity} of a NN allows us to understand the NN's theoretical potential for faithfully predicting such complex behavior. When the activation function of the NN is piecewise linear, such as with the ReLU activation function, then the output $\tilde{u}$ is a piecewise linear function of $(x,t)$, and the expressivity of an NN in this case is often measured as the number of $(x,t)$ connected geometric regions over which $\tilde{u}$ is linear. 

Consider a fully connected NN with $L$ hidden layers each of width $W$. Many recent studies have established that it is possible for the number of linear regions in such an NN to grow proportionally to $W^L$ \cite{pascanu_number_2014,montufar_number_2014}. I.e., growth is exponential in the depth and algebraic in the width. This implies that the NN can theoretically express functions of exponentially growing complexity as the number of layers is increased, or of algebraically growing complexity by increasing width. Futher refinements of this theme have gained traction in the NN literature, including results that articulate expressive limitations with increasing depth \cite{raghu_expressive_2017,telgarsky_benefits_2016,telgarsky_representation_2015,serra_bounding_2018}.

\subsection{A multifidelity method for Physics-Informed Neural Networks}
The theoretical expressivity of NN's provides one way to understand an accuracy-efficiency tradeoff with PINNs: NN architectures with larger width and/or depth have the potential to capture more complex PDE solutions, but this larger architecture is also more expensive to train through backpropagation and can require larger $N_R$ and $N_u$ in \eqref{eq:MSE}. Viewed in this light, a natural way to define fidelity for PINNs is through the network architecture. High-fidelity PINNs are those that have large width and/or depth, and low-fidelity PINNs have smaller width and depth.

The above interpretation of the accuracy-efficiency tradeoff as a notion of fidelity is based on theoretical insights about expressivity, but the practical realities surrounding expressivity are more complex: There is a large gap between the theoretical expressive potential of NN's and the realized expressivity of trained NN's in a practical setting. In particular, the exponentially expressive potential of increasing-depth networks is not frequently observed in practice, and with appropriate probabilistic models for weights of the network the average expressivity behaves linearly with respect to network depth \cite{hanin_complexity_2019}.

Acknowledging that a practical training procedure that ensures monotonic accuracy with respect to NN width/depth and/or optimization criteria is not yet readily available, we posit that PINNs that are \textit{low-fidelity} have simpler (less expressive) network architectures, and have looser optimization criteria. In contrast, \textit{high-fidelity} PINNs have more complex architectures (higher expressivity) and stricter optimization criteria. 

In this way, we can form a bi-level fidelity hierarchy with PINNs emulators. Let $\tilde{u}_L(x, t; \underbar{w}_L)$ denote a trained low-fidelity PINN, and let $\tilde{u}_H(x,t;\underbar{w}_H)$ denote a trained high-fidelity PINN. Here $\underbar{w}_L$ and $\underbar{w}_H$ denote the weight vectors associated to the high- and low-fidelity network architectures and optimization criteria and as such $\dim \underbar{w}_L \ll \dim \underbar{w}_H$. Since $\tilde{u}_L$ is less expressive than $\tilde{u}_H$, it should require less training time but also suffers from limited predictive power. The low-fidelity PINN $\tilde{u}_L$ is also trained with looser optimization criteria than $\tilde{u}_H$. Despite the practical caveat about tying architecture and optimization criteria to PINNs accuracy, we observe that increasing width and depth and strengthening optimization criteria does deliver a more accurate PINNs emulator, see Table \ref{table:performance} in Section \ref{sec:performance}. The particular choices of our network width, depth, and optimization criteria are provided in sections \ref{sssec:algorithm-complexity} and \ref{sec:results}.

In the parametric multifidelity problem of Section \ref{sec:multifidelity}, in order to construct a PINN for every parameter in the discretized parameter space $D_L = \{p_1, \ldots, p_K\}$, we must train $K$ PINN's solutions. This is perhaps feasible for $\tilde{u}_L$, whose architecture is simpler and optimization criteria are looser, but not for the more expensive $u_H$. To address this difficulty, we employ a multifidelity procedure.

Let $\Xi \subset \Omega \times [0, T]$ denote the spatiotemporal domain of the PDE. (We present the methodology in the general spatiotemporal case; for a stationary PDE problem, we set $\Xi = \Omega$.) We let $\widetilde{\Xi}$ be a set of test points for the PINN, i.e., a size-$P$ discretization of $\Omega \times [0, T]$ that are distinct from the PINN training set (i.e. collocation points) in this paper. Then the parameter models $g_L$ and $g_H$ are defined as the trained PINN at parameter values $p$ evaluated on the grid $\widetilde{\Xi}$,
\begin{linenomath}\begin{align}
  g_L(p) &\coloneqq \left( \tilde{u}_L(x_i, t_i; \underbar{w}_L(p) ) \right)_{i=1}^P \in \R^P, & 
  g_H(p) &\coloneqq \left( \tilde{u}_H(x_i, t_i; \underbar{w}_H(p) ) \right)_{i=1}^P \in \R^P,  
\end{align}\end{linenomath}
where $\underbar{w}_L(p)$ and $\underbar{w}_H(p)$ are the weights computed from training the low- and high-fidelity PINN, respectively, and the points $(x_i, t_i)$ are the spatiotemporal points that comprise $\widetilde{\Xi}$,
\begin{linenomath}\begin{align}\label{eq:Xi-def}
  \widetilde{\Xi} &= \left\{ (x_i, t_i) \right\}_{i=1}^P.
\end{align}\end{linenomath}
Note that in this context, $g_L(p)$ and $g_H(p)$ are both vectors in $P$-dimensional space, so that in the notation of Section \ref{sec:multifidelity} we have $m = M = P$. While one could choose different test grids for $u_L$ and $u_H$, we typically take $\widetilde{\Xi}$ to be a relatively dense and equispaced grid so that with an appropriate normalization the linear kernel/inner product produces an approximation to a continuous $L^2$ inner product, 
\begin{linenomath}\begin{align}\label{eq:discrete-continuous-ip}
  \left\langle g_L(p), g_L(q) \right\rangle \approx \int_{\Omega \times [0, T]} \tilde{u}_L(x,t;\underbar{w}_L(p)) \tilde{u}_L(x,t;\underbar{w}_L(q)) dx dt,
\end{align}\end{linenomath}
and similarly for the high-fidelity model $g_H$.

Given this multifidelity surrogate on weights in high-fidelity space, we generate a multifidelity PINN (which we call mfPINN) with high-fidelity expressivity from the procedure in Section \ref{sec:multifidelity}:
\begin{linenomath}\begin{align}\label{eq:PINN-multifidelity-relation}
  \widehat{u}_H(\widetilde{\Xi};p) \coloneqq \widetilde{g}_H(p), %u\left(x_i,t_i,\widetilde{g}_H(p); \ell_H, n_{1,H}, \ldots, n_{\ell,H}\right).
\end{align}\end{linenomath}
which is an approximation to $\tilde{u}_H(\widetilde{\Xi}; \underbar{w}_H(p))$ that is produced without the need to train $\underbar{w}_H(p)$. We recall that computing $\widetilde{g}_H(p)$ requires the single low-fidelity evaluation $g_L(p)$, i.e., a single training of $\tilde{u}_L$ at parameter $p$.
If the cost of training $\tilde{u}_H$ is much larger than the cost of training $\tilde{u}_L$, then an accurate surrogate $\widehat{u}_H(p)$ can therefore be delivered with the training cost of only $\tilde{u}_L$.

\begin{rmk}
The current state-of-the-art theory in DNNs provides {\em a priori} consistency statements about PINN solutions.  However, a complete convergence theory remains as a topic of future work.  Correspondingly, the results shown in this paper do not imply that every increased combination of depth and/or width of the network leads to improved accuracy of the DNN solution.  The interplay of fidelity (expressivity) and optimization choices is a current area of research.
\end{rmk}

%%%%%%%%%%%%%%%%%%%%%%%%%%%%%%
\subsubsection{Theoretical Considerations}\label{sssec:theory}

Theoretical analysis for the particular low-rank multifidelity surrogate of Section \ref{sec:multifidelity} is developed in \cite{narayan2014stochastic,keshavarzzadeh19SIAM,hampton2018practical}. The core requirements are that (i) the manifold of vectors $\left\{g_H(p) \;\; \big|\;\; p \in D \right\}$ is low-rank, and (ii) that one has an inner product proximity statement of the form,
\begin{linenomath}\begin{align}\label{eq:ip-proximity}
  \left\langle g_L(p), g_L(q) \right\rangle \approx \left\langle g_H(p), g_H(q) \right\rangle,
\end{align}\end{linenomath}
where $\left\langle \cdot, \cdot \right\rangle$ is the standard Euclidean inner product on vectors as in \eqref{eq:E6} and $p$ and $q$ denote positions within the sampled parameter space. In particular, it is \textit{not} necessary that $g_L(p)$ be a good approximation to $g_H(p)$. Instead, one requires similar \textit{parametric} dependence in the maps $p \mapsto g_L(p)$ and $p \mapsto g_H(p)$. We recall that with the dense grid $\widetilde{\Xi}$, then the inner product acting on evaluations of $g_L$ approximates the continuous $L^2(\Omega \times [0, T])$ inner product, cf. \eqref{eq:discrete-continuous-ip}. Under the assumption that $\tilde{u}_L$ and $\tilde{u}_H$ are approximations of appropriate accuracy to the true PDE solution $u$, then analysis can be carried out by exploiting \eqref{eq:ip-proximity} and the relationship to continuous $L^2$ norms \eqref{eq:discrete-continuous-ip} to conclude that the multifidelity procedure converges. For example, such an analysis exploits small Kolmogorov $n$ width of the manifold induced by $g_H$ \cite{narayan2014stochastic}, and can utilize accuracy estimates of $\tilde{u}_L$ and $\tilde{u}_H$ relative to $u$ \cite{keshavarzzadeh19SIAM}. Many of the previously mentioned analysis techniques are theoretical in nature and can be difficult to apply in complex settings. Alternatively, the approach in \cite{hampton2018practical} provides a computational strategy that produces model-independent algorithmic bounds on quantities similar to \eqref{eq:ip-proximity} that lead to estimates for error of $\widetilde{g}_H$.

%%%%%%%%%%%%%%%%%%%%%%%%%%%%%%
\subsubsection{Implementation Considerations}

The multifidelity approach laid out in Section \ref{sec:multifidelity} contains four steps, summarized as:  (1) sample the parameter space; (2) evaluate the low-fidelity model at 
the aforementioned samples and decide at which locations one should run the more costly high-fidelity model; (3) evaluate the high-fidelity model on the subset of ``important" points; and
lastly (4) use the set of low- and high-fidelity simulations to construct a multifidelity emulator.  This multifidelity emulator, when queried at a new (not previously evaluated) location within the 
parameter space requires the evaluation of only the low-fidelity model and the new location combined with a computationally efficient manner of augmentation (which is significantly 
cheaper than evaluation of the high-fidelity model outright) to provide an updated fidelity response.  

Based upon the theoretical considerations presented above,  only Step (2) in the multifidelity procedure requires systematic modification.  Implicit in Step (2) is the decision as to what
fidelity means in the context of this approach and how is it tuned.  In the case of PINNs, we have selected width and depth of the neural network architecture (with fixed activation functions) 
as the our adaptable fidelity hyperparameters.

%%%%%%%%%%%%%%%%%%%%%%%%%%%%%%
\subsubsection{Algorithmic Complexity}\label{sssec:algorithm-complexity}
To aid in evaluating the algorithmic complexity and reproducibility of our approach, we provide Algorithm \ref{mf-algo}. Note that the first three steps: storing the $p^C$ weights, storing trained low-fidelity solutions, and storing trained high-fidelity solutions, are offline steps. The online step is generating a new low-fidelity solution and computing the coefficients in Equation \mbox{\ref{eq:E9}} for the multifidelity approximation. For this algorithm we define the notation for the centroid of domain $D$ as $p^C$. Given the parametric PDE domain bounds $[a_i,b_i]$ in dimension $i$
\begin{linenomath}\begin{align*}
  p^C =  \frac{a_i+b_i}{2}, \qquad i = 1,...,d
\end{align*}\end{linenomath}

The main considerations when evaluating the cost of our approach are the sizes $K$ and $k$ of the low and high fidelity sample sets respectively, and the computational costs of evaluating $g_L$ and $g_H$ associated with a PDE. Given that it is often the case that cost of $g_H$ is much greater than the cost of $g_L$,  minimizing $k$ provides the greatest cost savings in our approach. 

\begin{algorithm}
\caption{Multifidelity PINNs Method} 
\label{alg1}
\begin{algorithmic}
    \STATE Given appropriate set-up information (i.e. domain discretization $\widetilde{\Xi}$, low-fidelity parameter space sampling $D_L$, PDE initial and boundary conditions, low \& high-fidelity sample sizes $K$ \& $k$, PINN hyper-parameters (size and optimization) for $\tilde{u}_L$ \& $\tilde{u}_H$  etc.)
    \STATE Assume $\frac{||u-\tilde{u}_L||_2}{||u||_2} > \frac{||u-\tilde{u}_H||_2}{||u||_2}$
    \STATE Assume $\tilde{u}_L \: training \: time < \tilde{u}_H \: training \: time$
    \STATE Store trained weights $\underline{w}_L(p^C)$ from $\tilde{u}_L(\widetilde{\Xi};\underline{w}_L(p^C))$
    \FOR {n = 1 \TO $K$}
    	\STATE Initialize $\tilde{u}_L \: $weights$ \: \underline{w}_L(p_n)$ with $\underline{w}_L(p^C)$
    	\STATE Store trained $\tilde{u}_L(\widetilde{\Xi};\underline{w}_L(p_n))$ solutions
    \ENDFOR
    \STATE Compute Gram matrix $G_L$ as defined in \eqref{eq:E6}
    \STATE Set $D_H$ by selecting $k$ points in $D_L$ identified from the first $k$ pivots in the pivoted Cholesky decomposition of $G_L$ as described in Section 2
    \STATE Store trained weights $\underline{w}_H(p^C)$ from $\tilde{u}_H(\widetilde{\Xi};\underline{w}_H(p^C))$
    \FOR {n = 1 \TO $k$}
        	\STATE Initialize $\tilde{u}_H \: $weights$ \: \underline{w}_H(p_{i_n})$ with $\underline{w}_H(p^C)$
    	\STATE Store trained $\tilde{u}_H(\widetilde{\Xi};\underline{w}_H(p_{i_n}))$ solutions
    \ENDFOR
    \STATE To use multifidelity method: solve $\tilde{u}_L$ at any point $p$, then perform the multifidelity procedure in \eqref{eq:mf-surrogate} \& \eqref{eq:E9} using the stored $\tilde{u}_L$ \& $\tilde{u}_H$ solutions to construct $\widehat{u}_H(\widetilde{\Xi};p)$ which emulates $\tilde{u}_H(\widetilde{\Xi};\underline{w}_H(p))$
\end{algorithmic}
\label{mf-algo}
\end{algorithm}

\begin{rmk}\label{rmk:1}
The optimization process for DNNs is highly dependent on the initial state of the weight matrix that is used \cite{DeepLearning}.   The non-convex nature of the optimization problem often leads to many local minima, all of which have very similar loss function evaluations.  In some applications like metalearning of PINNs, smooth transitions in the weight matrices across parameterized runs are desired \cite{Penwarden2021a}.  In this work, we have initialized our weight matrices by first computing a PINN solution at the midpoint of the parametric domain and then using that weight matrix as initialization for all subsequent parametric point evaluations. This is known to reduce not only cost but also reduce the likelihood that a PINN will not be trained well, which is important in the online step.
\end{rmk}

In regards to optimization, all runs (both low and high) are initially optimized with $500$ epochs using the Adam version of stochastic gradient decent (SGD) with a learning rate of $10^{-3}$. This is to help get in the vicinity of a loss function minimum before using the more precise L-BFGS optimizer, which has been shown to aid the consistency of optimizing DNNs as well as PINNs. Without doing employing this strategy, we encountered more ``bad" runs where using \textit{only} L-BFGS does not optimize well. 

In the context of optimization criteria and in terms of L-BFGS, we specify that for low-fidelity runs the maximal number of iterations per optimization step is $5000$, the termination tolerance on a first order optimality condition is $10^{-6}$, and the termination tolerance on weight vector changes is $10^{-9}$. For high-fidelity runs, these are $10,000$, $10^{-9}$, and $10^{-12}$ respectively. We found that this helps increase the accuracy and time differences between low and high-fidelity which give more distinct results. (We again emphasize that PINNs, and DNNs in general, do not in practice always yield higher accuracy with a more expressive architecture even with a very large amount of data.)  As mentioned earlier, as a pre-optimization step, we initialize the weights of the PINN with the final weights of a PINN run at the center of the PDE hyper-parameter range for low and high respectively.

\section{Results and Discussion}
\label{sec:results}

In this section, we demonstrate the efficacy of our multifidelty approach on four forward PDE problems in one and two spatial dimensions: 1D Burgers' equation, a 1D nonlinear heat equation, the 2D nonlinear Allen-Cahn equation, and a 2D nonlinear Diffusion-Reaction equation.  These examples are extensions of the test problems proposed in \cite{raissi2019physics,Yang2020}.
For all the PINNs architectures used below, we employ fully-connected feed-forward neural networks with tanh
activation functions. There are two types of snapshots spaces here: (i) the Burgers' and (nonlinear) Heat equations in which we have a spatial and temporal dimension, and the (ii) Allen-Cahn and Diffusion-Reaction equations in which we have two spatial dimensions. For the former, the test set of points $\widetilde{\Xi}$ is a uniform grid of size $256 \times 100$ in space and time respectively. For the latter we use a uniform grid of size $128 \times 128$. These test points are where we compare between the exact solution and the PINN solution. To train the PINN we use $100$ uniformly sampled boundary/initial value points for the $MSE_u$ portion of the loss, and $10,000$ collocation points using Latin hypercube sampling (LHS) for the $MSE_R$ portion as described in Section \ref{sec:pinns}.

%%%%%%%%%%%%%%%%%%%%%%%%%%%%%%%%%%
\subsection{1D Burgers Equation}

We consider the following 1D viscous Burgers equation:

\begin{equation}
\frac{\partial u}{\partial t} + \frac{1}{2}\frac{\partial}{\partial x} \left( u^2 \right) = \nu \frac{\partial^2 u}{\partial x^2} 
\label{eq:burgers}
\end{equation}

\noindent on $(x,t) \in \Xi =  [-1,1] \times [0,1]$ with the viscosity $\nu \in [0.005,0.05]$ and initial condition $u(x,0) = -\sin(x\pi)$. Our parameter in this example is the viscosity, i.e., $p = \nu$.
For evaluation of the error, we compute the exact solution derived using Cole's transformation computed with Hermite 
integration \cite{Patera86}. Our sampled parameter space $D_L$ is constructed with $K=50$ LHS points, and the low-fidelity PINN $\tilde{u}_L$ has 
$2$ hidden layers each of width $5$. 
Based upon our multifidelity procedure, we evaluate $k=10$ 
high-fidelity PINNs samples, chosen by the pivoted Cholesky decomposition of the low-fidelity snapshot matrix. The high-fidelity PINN $\tilde{u}_H$ consists of $5$ hidden layers of width $10$. Using the collected low- and high-fidelity ensembles, we can perform the multifidelity procedure on any future low fidelity runs to evaluate the multifidelity emulator $\widehat{u}_H$. For this case with one PDE parameter in the Burgers' equation, we generate test points over $100$ uniformly sampled points and compare to the exact solution to generate the plot in Figure \ref{eq:burgers}.

\begin{figure}[H]
\centering
\includegraphics[scale=0.7]{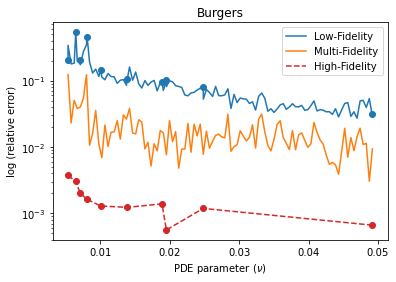}
  \caption{Line plot of the log (base $10$) relative error between the exact solution and PINN solution at a given parameter, in this case viscosity as defined in Equation \ref{eq:burgers}. Note that the multifidelity emulator error is sandwiched between the low and high fidelity runs, as would be expected. This behavior is not guaranteed if the high-fidelity PINNs are inaccurate compared to the low-fidelity ones, but this depends on the formulation of the problem attempted. If it is a well posed problem with well chosen parameter ranges, low and high-fidelity DNN sizes, etc. then this behavior is frequently observed.}\label{fig:burgers}
\end{figure}

The results in Figure \ref{fig:burgers} show that low fidelity accuracy can be enhanced with this multifidelity procedure, if one invests in training a modest number of higher-fidelity PINNs. We also observe that the relative error decreases with increasing viscosity, which aligns with expectations that small viscosity corresponds to regimes where shocks can form, which are generally more difficult to approximate. The red and blue dots represent the $k=10$ important parameters locations selected by the multifidelity procedure. We observe that the pivoted Cholesky decomposition clusters points toward lower viscosity, which again agrees with expectations since in that near-shock regime the parametric variations are more complex. Lastly, the noisy fluctuations can be explained since DNNs have substantial variance in their solutions due to randomness imparted during training (e.g., with Adam), so the accuracy of these solutions does not vary smoothly with PDE parameter. We attempt to reduce these fluctuations using previously described methods, such as running Adam before L-BFGS and initializing the weights using a previous PINN solution at the parameters center. 

%%%%%%%%%%%%%%%%%%%%%%%%%%%%%%%%%%
\subsection{1D nonlinear Heat Equation}

We consider the following 1D nonlinear PDE:

\begin{equation}
\frac{\partial u}{\partial t} - \lambda \frac{\partial^2 u}{\partial x^2} + k \,\tanh(u) = f, \, x\in\Omega
\label{eq:1dheat}
\end{equation}

\noindent where $(x,t) \in \Xi = [-1,1] \times [0,1]$ and where $\lambda \in [1,\pi]$ and $k\in[1,\pi]$ are positive constants.  Our parameter in this example will be the joint tuple $p = (\lambda, k)$.
In order to compute errors, we employ the method of manufactured solutions, specifying an exact solution of $u(x,t;\lambda,k) = k\,sin(
\pi x)\, exp(-\lambda k x^2)\,exp(-\lambda t^2)$ and
derive the corresponding form of the forcing $f$. We choose $D_L$ as $K=50$ LHS samples in parameter space. The low-fidelity PINNs have $2$ hidden layers each of width $5$. Based upon our multifidelity procedure, we choose $k = 10$ important parameter locations and evaluate that many high-fidelity PINNs. The high-fidelity PINN has an architecture of $5$ hidden layers of width $10$. For this case with two-dimensional parameter space, we use an equidistant $20 \times 20$ grid as the set of test points.

\begin{figure}[H]
\centering
\includegraphics[scale=0.5]{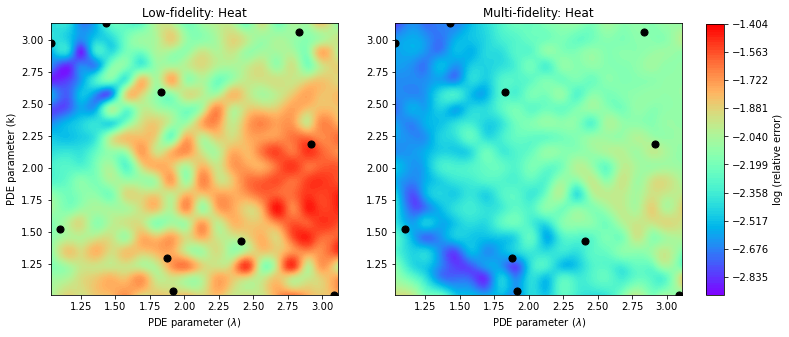}
  \caption{Surface plot of the log (base $10$) relative errors between the exact solution and PINN solution at given parameters, in this case $k$ and $\lambda$ as defined in Equation 13. The black dots represent the parameter locations at which the multifidelity procedure was constructed with $k=10$ high fidelity runs. }\label{fig:heat}
\end{figure}

Results are shown in Figure \ref{fig:heat}, where the shown surface plots correspond to a $100 \times 100$ grid that is generated via cubic spline interpolation from the $20 \times 20$ test grid. Note that the cubic spline procedure visually smooths out noisy fluctuations, cf. Figure \ref{fig:burgers}. The multifidelity procedure clearly improves accuracy in the parameter region chosen, and we can see the general trend is that low $k$ values and high $\lambda$ values have higher errors for the low-fidelity PINN. We observe that the multifidelity procedure ameliorates this inaccuracy.

%%%%%%%%%%%%%%%%%%%%%%%%%%%%%%%%%%
\subsection{2D nonlinear Allen-Cahn Equation}

We consider the following 2D nonlinear Allen-Cahn equation, which is a widely used model for multi-phase flows:

\begin{equation}\label{eq:allen-cahn}
\lambda \left( \frac{\partial^2 u}{\partial x^2} + \frac{\partial^2 u}{\partial y^2}\right) + u\left(u^2-1\right) = f, \, x,y\in\Omega
\end{equation}

\noindent where $\Xi = [-1,1]^2$,
with $\lambda \in (0,\pi]$ the mobility, and $u$ denotes the order parameter, prescribing different phases.  The parameter in this problem is $p = \lambda$.
We again use the method of manufactured solutions, specifying an exact solution of $u(x,y;\lambda) = \exp(-\lambda (x+0.7)) \,\sin(\pi x) \,\sin(\pi y)$ and derive the corresponding form of the forcing $f$. Parameter space is discretized with $K=50$ sampled uniformly at random.
The low-fidelity PINN has $5$ hidden layers of width $5$, and the high-fidelity PINN has $8$ hidden layers of width $10$. We budget $k=10$ high fidelity PINNs evaluations.

\begin{rmk}
Both the low and high fidelity architectures are more complex and expressive than the ones used in the previous Burgers' and Heat equation examples.  It is likely that the nature of the PDE being solved will strongly influence what sizes are appropriate for each fidelity.  These architectural parameters were manually tuned to give reasonable results, and automatic procedures for such hyperparameter selection is the topic of ongoing investigations.
\end{rmk}

\begin{figure}[H]
\centering
\includegraphics[scale=0.7]{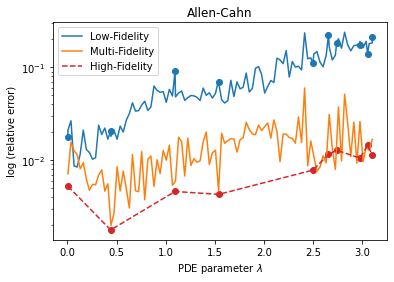}
  \caption{Log (base $10$) relative error between the exact solution and PINN solution at a given parameter, in this case $\lambda$ as defined in \eqref{eq:allen-cahn}.}
\end{figure}

With this Allen-Cahn example, we observe the inverse relation between error and PDE parameter to that in Burgers. Here as $\lambda$ increases, so does the solution complexity and therefore the error. We also observe the clustering of multifidelity ``important points" in the region with the highest error.

\begin{rmk}
Note that around $\lambda = 2.7$, we observe that the multifidelity approach has a lower error than the high fidelity points and considerable oscillations over the domain. The error of the multifidelity procedure is not strictly bounded by the low and high fidelity methods: There is a considerable variance when training neural networks as the process itself is randomized in the initialization and/or optimization. Each time a network is trained, the results may be slightly different even at the same parameter locations. This stochasticity implies that we cannot ensure that the multifidelity error is strictly bounded by the low- and high-fidelity errors. The oscillations/variance observed are also a direct result of this stochasticity.
\end{rmk}
%%%%%%%%%%%%%%%%%%%%%%%%%%%%%%%%%%
\subsection{2D nonlinear Diffusion-Reaction Equation}

\begin{figure}[H]
\centering
\includegraphics[scale=0.5]{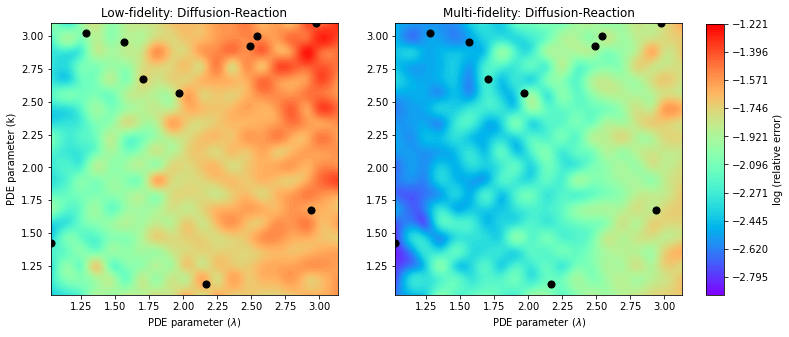}
  \caption{Surface plot of the log (base $10$) relative error between the exact solution and PINN solution. The parameters are $k$ and $\lambda$ as defined in \eqref{eqn:diffusion-reaction}. The black dots show the $k=10$ important points in parameter space selected by the pivoted Cholesky procedure.}\label{fig:diffreac}
\end{figure}

We consider the following 2D nonlinear diffusion-reaction equation:
\begin{equation}
\lambda \left( \frac{\partial^2 u}{\partial x^2} + \frac{\partial^2 u}{\partial y^2}\right) + k \left(u^2\right) = f, \, x,y\in\Omega
\label{eqn:diffusion-reaction}
\end{equation}
\noindent where $(x,y) \in \Xi = [-1,1]^2$.  
Here $\lambda \in [1,\pi]$ represents the diffusion coefficient and $k\in[1,\pi]$ represents the reaction rate and $f$ denotes the source term. Our parameter $p$ is the tuple $(\lambda, k)$.
We specify an exact solution as $u(x,y;\lambda,k) = k\, \sin(\pi x)\, \sin(\pi y) \,\exp(-\lambda \sqrt{(k\,x^2+y^2)})$ 
and derive the corresponding form of the forcing $f$. The low-fidelity PINN has $5$ hidden layers of width $10$, and the high-fidelity PINN has $8$ hidden layers of width $20$. Parameter space is discretized with $K=50$ points drawn uniformly at random. Results of this experiment are shown in Figure \ref{fig:diffreac}.
Similar to the Heat equation example, we observe that the multifidelity procedure improves accuracy compared to the low-fidelity PINN.

%%%%%%%%%%%%%%%%%%%%%%%%%%%%%%%%%%
\subsection{Performance Metrics}
\label{sec:performance}
We summarize numerical accuracy and cost measurements of the low-fidelity and high-fidelity PINNs, and of the multifidelity emulator in Table \ref{tab:title}.  These experiments were run on an Intel Core i7-5930K processor with the Windows 10 OS. PyTorch version 1.6.0 and Python version 3.8.5 were used to run the PINNs. Average values and standard deviations of cost and error computed over parameter space are reported.
\begin{center}
  \captionof{table}{Metrics of performance taken for each of the four problems where $\pm$ indicates 1 standard deviation. For relative error the values presented are the mean of the PDE parameter 1D or 2D grids sampled as shown in the previous subsections. The computational time (in seconds) corresponds to training time in the L-BFGS portion of the optimization for the indicated PINNs.} \label{tab:title}
\scalebox{0.75}{
\begin{tabular}{| c | c | c | c | c |}
\hline
PDE's & 1D Burgers & 1D Heat & 2D Allen-Cahn & 2D Diffusion-Reaction \\
\hline
Low-Fidelity Error $(10^{-2})$ & $7.86 \pm 5.77$ & $1.24 \pm 0.69$ & $7.71 \pm 5.61$ & $1.82 \pm 0.91$ \\

Multi-Fidelity Error $(10^{-2})$ & $1.82 \pm 1.76$ & $0.55 \pm 0.25$ & $1.39 \pm 0.92$ & $0.81 \pm 0.51$ \\

High-Fidelity Error $(10^{-2})$ & $0.17 \pm 0.09$ & $0.46 \pm 0.25$ & $0.85 \pm 0.41$ & $0.75 \pm 0.51$ \\

Low-Fidelity Time (s) & $41 \pm 19$ & $63 \pm 12$ & $167 \pm 32$ & $212 \pm 13$ \\

High-Fidelity Time (s) & $137 \pm 72$ & $181 \pm 64$ & $403 \pm 137$ & $890 \pm 207$ \\
\hline
\end{tabular}}
\label{table:performance}
\end{center}

The low and high-fidelity PINNs correspond to less and more accurate solvers, respectively, and also correspond to less and more costly solvers, respectively. Since the multifidelity emulator cost equals that of the low-fidelity PINN, this demonstrates significant potential for savings in parametric multi-query contexts. The high-fidelity solver is around twice as expensive for the Allen-Cahn example, and around $4$ times as costly for the Diffusion-Reaction example. 
In every test problem, the mean error of the multifidelity offers substantial improvement over its low-fidelity counterpart. 

%%%%%%%%%%%%%%%%%%%%%%%%%%%%%%%%%%
\subsection{Convergence Plots}
\label{sec:convergence}

\begin{figure}[H]
\centering
\includegraphics[scale=0.5]{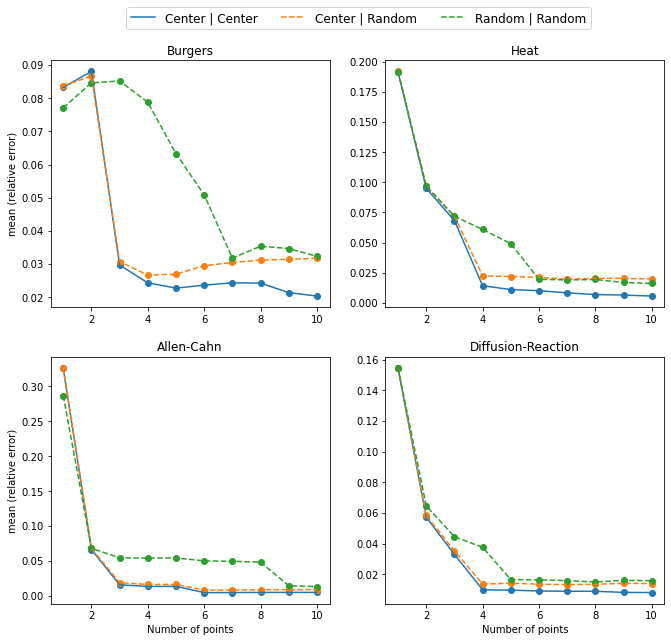}
  \caption{Convergence plots of the four PDE problems: Burgers, Heat, Allen-Cahn, Diffusion-Reaction. Three cases were run: one in which we initialize with the center of the parameter space for both training and testing (blue), one in which we initialize for training but not testing (orange), and one in which we use standard randomized weights for both steps (green). The multifidelity method converges regardless of this choice, but it is ideal to initialize for both steps, as seen by blue line}
 \label{fig:convergence}
\end{figure}

We provide Figure \mbox{\ref{fig:convergence}} to show the convergence of the example problems as more points are added in the multifidelity method, giving more insight into the accuracy-cost tradeoffs of our method. Our results in previous sections utilized a procedure corresponding to the blue line in Figure \mbox{\ref{fig:convergence}}. In this figure, the labels ``center" and ``random" are in reference to the NN weight initialization; the first word refers to the initialization used for constructing the multifidelity method, and the second for generating the test points. E.g., ``Center | Random" denotes that the multifidelity initialization is chosen according to the \textit{center} run and that the error on test points is computed by constructing a PINN whose initialization is \textit{random}. As briefly discussed in Remark \mbox{\ref{rmk:1}} in Section \mbox{\ref{sssec:algorithm-complexity}}, we initialize with the trained weights at $p^C$. For additional intuition about this, we refer the reader to \mbox{\citep{Penwarden2021a}}. Using this strategy, we not only observe a decrease in cost, but as seen in the convergence plots for all problems, the multifdelity method converges as well. By initializing in this way, we empirically obtain smooth NN weights in the parametric domain. It is then logical that in our multifidelity method, we observe better behavior due to less variance in the weights and, therefore, in the solution. However, we provide an experiment corresponding to the green line in Figure \mbox{\ref{fig:convergence}}, where all initialization is random, to demonstrate that even with standard initialization approaches our method works well. We also show the orange line in Figure \mbox{\ref{fig:convergence}}, showing a center initialization in training but a random initialization during the test phase, to demonstrate efficacy even if training uses a center initialization, but a different (random) initialization is used in the test phase. Finally, our method is shown to converge after only a few points even though we provide solutions with ten points so one could cut down on offline cost even more in practice.

\section{Summary and Conclusions}
\label{sec:summary}

In this paper, we have extended the low-rank multifidelity approach of \cite{narayan2014stochastic,zhu2014computational} to Physics-informed Neural Networks (PINNs) \cite{raissi2019physics,raissi2017physicsI,raissi2017physicsII}. Having summarized the multifidelity approach and the collocation version of PINNs, we construct a low-fidelity version of a PINN as one with a simpler (less expressive) architecture and less stringent optimization termination criteria. We empirically demonstrate that low-fidelity PINNs constructed in this way can be both less costly and less accurate than high-fidelity versions. This motivates our proposed multifidelity approach, whose computational results on several PDEs demonstrate that the multifidelity emulator can provide an accuracy-increasing PINNs surrogate over a PDE-based parameter space at significant savings in computational cost. We also observe that initializing the weights with a \textit{center} run as in \mbox{\citep{Penwarden2021a}} helps to speed up convergence as a function of multifidelity points used.

Future investigations will explore more quantitative connections between PINNs architecture and accuracy/cost tradeoffs. In addition, the multifidelity approach considered here is agnostic to the type of PDE solver employed. We have used PINNs as both low and high-fidelity models. However, for example, one could utilize PINNs for a low-fidelity model and more traditional numerical solvers (such as finite element methods) for the high-fidelity model. Such combinations could generate a multifidelity surrogate that combines the advantages of different types of solvers.

\vspace{0.2in}
\noindent {\bf Acknowledgements:}
The authors would like to acknowledge helpful discussions with Professor George Karniadakis and his group (Brown University).  This work was funded under AFOSR MURI FA9550-20-1-0358. A.~Narayan was partialy supported by AFOSR FA9550-20-1-0338.

\newpage
\noindent {\bf Bibliography}
\bibliography{references}

\end{document}